\documentstyle[aps,prl,multicol]{revtex}

\begin{document}
\draft
\preprint{}
\title{Spontaneous magnetic moments in
YBa$_2$Cu$_3$O$_{7-\delta}$ thin films}
\author{F. Tafuri}
\address{Dip. di Ingegneria dell'Informazione, Seconda Universit\'{a} di Napoli, Aversa (CE) and\\
INFM - Dip. di Scienze Fisiche, Universit\'{a} di Napoli Federico II, Napoli, Italy}
\author{J.R. Kirtley}
\address{ IBM T.J. Watson Research Center, P.O. Box 218, Yorktown Heights, NY 10598, USA}
\date{\today}
\maketitle
\begin{abstract}
We have observed spontaneous magnetic moments with random signs
in $c$-axis oriented thin films of the high-T$_c$ cuprate superconductor
YBa$_2$Cu$_3$O$_{7-\delta}$ (YBCO),
imaged with a scanning SQUID microscope. These
moments arise when the samples become superconducting,
and appear to be associated with non-ferromagnetic defects in the films.
In contrast with granular high-T$_c$ samples, which also show spontaneous
moments with random signs, the present samples shield diamagnetically.

\end{abstract}
\begin{multicols}{2}
\narrowtext

\pagebreak
\narrowtext
It now appears likely that the superconductivity in optimally doped
cuprate high-T$_c$ superconductors can be characterized by
an orbital component of the pairing wavefunction with
predominantly $d_{x^2-y^2}$ symmetry\cite{scalrev,annett,trirmp}.
The $d_{x^2-y^2}$ order parameter does not violate time reversal
symmetry\cite{bailey,sigrist}, and
little, if any, time reversal symmetry breaking is observed
experimentally in the bulk material\cite{spielman}.
Nevertheless, it has been shown theoretically that broken time-reversal symmetry
(BTRS) could occur locally in a $d_{x^2-y^2}$ superconductor
at certain surfaces and interfaces,
or in the presence of non-magnetic and magnetic
impurities\cite{matsumoto,kuklov,simon,balatsky,zhu}.
The existence of a lowest
energy state which is not unique (BTRS) requires an additional component
of the order parameter that is $\pi/2$ out of phase with the
$d_{x^2-y^2}$ component. A $d_{x^2-y^2}+is$ or $d_{x^2-y^2}+id_{xy}$
pairing state has been suggested near a surface or twin boundary with
(110) orientation, producing a local nodeless global gap
function\cite{balatsky,zhu,salkola}. The predicted BTRS should manifest
itself in many effects, including deviations from the well-established
Josephson's sinusoidal current-phase relationship \cite{il'ichev}, and
the splitting of the zero bias conductance peak induced by the Andreev
surface bound states \cite{covington}. An intrinsic phase shift in
Josephson junctions is a signature of BTRS associated with a grain boundary
or a barrier interface and should produce
spontaneous supercurrents, possible fractional flux quanta and phase
slips \cite{bailey}.

However,
experimental measurements using scanning SQUID microscopy have shown no
evidence for BTRS in YBCO tricrystal samples
\cite{spinnat,vartsci}, in YBCO-Pb thin-film SQUIDs \cite{mathai}, or in
sparsely twinned YBCO single crystals \cite{kamfrac}. These samples each have
some combination of
grain boundaries, tunnel junctions, and twin boundaries, which might be
expected to induce $d_{x^2-y^2}+is$ or $d_{x^2-y^2}+id_{xy}$ local
pairing in a bulk $d_{x^2-y^2}$ superconductor.
On the other hand, measurements of
the magnetic field dependence of the zero-bias anomaly in
thin film YBCO/I/Cu tunnel junctions \cite{covington}, and
a spontaneous net magnetic moment in
bulk SQUID magnetization measurements of $c$-axis YBCO thin films
\cite{polturak}, support BTRS.
We note that in these last two experiments, evidence
for broken time-reversal symmetry was not seen in all samples.
In this paper we provide experimental evidence for large spontaneous
flux generation in YBCO c-axis oriented thin films, apparently associated
with non-ferromagnetic defects in the films. Such spontaneous moments
could result in the effects reported in support of BTRS
previously \cite{covington,polturak}. We speculate on the source of
our observed magnetic moments.

Our measurements were made on biepitaxial 45$^o$ø a-axis tilt
(twist) grain boundary Josephson junctions (GBJs) employing films
with different orientations \cite{tafprb}
(Fig. 1(a) is, for example, the schematic of a tilt GBJ imaged in Fig. 1(b)).
Details on the scanning SQUID measurements
can be found elsewhere\cite{ssmapl}. Our
samples employed a (110) MgO film as a seed layer to modify the crystal
orientation of YBCO on the (110) SrTiO$_3$ substrate. YBCO grows predominantly
along the [103] or [013] directions on the SrTiO$_3$ substrate, and along the
[001] direction on the MgO seed
layer. These structures allow us to investigate
regions with different properties and morphologies simultaneously.
Details of the fabrication process of the junctions have been described
elsewhere, along with a complete characterization of the microstructure
and transport properties \cite{tafprb}.
These films are affected by the presence of Y-based
impurities, as is commonly observed in YBCO thin films deposited by
sputtering \cite{tafphysicac}.
Y$_2$O$_3$ and Y$_2$BaCuO$_5$ (Y211) precipitates are in general
found in the (103) and (001) electrodes, as shown by transmission
electron microscopy (TEM) analyses\cite{tafprb,tafphysicac}.
The amount
of the precipitates can be modified and very roughly controlled by changing
the YBCO deposition conditions. The presence of Y211 precipitates can be
revealed by Scanning Electron Microscopy (SEM) or Atomic Force
Microscopy (AFM) with some accuracy.
TEM analysis shows that there
are no grain boundaries in the (001) films \cite{tafprb,tafphysicac}.

The spatial distribution of the magnetic flux for
a (001)/(103) 45$^o$ twist biepitaxial grain boundary sample is shown
in Fig. 1(b). The sample was cooled
in a field of $\sim$5 mG  and imaged at 4.2K
with an octagonal shaped low-T$_c$ SQUID sensor pickup loop
4 microns in diameter.
In this image
the (103) YBCO film is magnetically
smooth, aside from a single dipole feature (often seen in YBCO films), and
4 vortices trapped in the bulk
of the film. The vortices in the (103) film are anisotropic, elongated
in the a-b plane direction.
In contrast, the (001) YBCO on the MgO seed layer has
many localized regions of flux,
with the fields pointing both out
of (white) and into (black) the sample.
When the sample is cooled in zero field, the (103)
section has no vortices, but the (001) section still shows
many localized magnetic moments, averaging to zero net magnetization.
The (001) region is so magnetically disordered that vortices
are not well separated, and therefore estimates of the total flux
in each ``vortex" are difficult. Nevertheless,
modelling of the most isolated regions of flux give
values for the total flux
significantly less than the superconducting flux quantum.
An example is shown in Fig. 1(c), which shows experimental data
and modelling for a cross-section through Fig. 1(b) as indicated
by the nearly horizontal dotted line. This cross-section
passes directly through two vortices on opposite sides of the
grain boundary (indicated by the nearly vertical dashed line).
We model the vortices in the (103) region
as anisotropic vortices emerging normal
to a surface and parallel to the planes of a layered superconductor.
The $z$-component of such a superconducting
vortex is given by \cite{kogprb}:
\begin{equation}
h_z({\bf r},z)=-\int \frac{d^2 {\bf k}}{(2\pi)^2} k \phi({\bf k})
e^{i{\bf k \cdot r}-kz},
\end{equation}
where
\begin{equation}
\phi({\bf k}) = - \frac{\phi_0 (1+m_1 k_x^2)}{m_3 \alpha_3[m_1 k_x^2 \alpha_3
(k+\alpha_1)+k\alpha_3+k_y^2]},
\end{equation}
$\alpha_1=((1+m_1k^2)/m_1)^{1/2}$, $\alpha_3=((1+m_1k_x^2+m_3k_y^2)/m_3)^{1/2}$,
$k=(k_x^2+k_y^2)^{1/2}$,
$m_1=\lambda_{ab}^2/\lambda^2$, $m_3=\lambda_c^2/\lambda^2$, $\lambda =
(\lambda_{ab}^2 \lambda_c)^{1/3}$, $\lambda_{ab}$
is the in-plane penetration depth,
$\phi_0 = hc/2e$ is the superconducting
flux quantum, $h$ is Planck's constant,
$e$ is the charge on the electron, $x$ is the distance
perpendicular to the planes, and $y$ is
the distance parallel to the planes. The $z$-component of the field is
integrated over the known geometry of the pickup loop to obtain a flux
to compare with experiment. The solid line through the vortex
to the right of Fig. 1(c) is a fit of Eq. 1 to the experiment (dots)
with the height $z$=2.45$\mu$m of the pickup loop and the
$c$-axis penetration depth $\lambda_c$=5.9$\mu$m as the two fitting
parameters, with $\lambda_{ab}$=0.14$\mu$m held fixed. For comparison,
the solid line
to the left of Fig. 1(c) is the expected flux through the pickup loop
for a monopole vortex field source with $\phi_0$ total flux,
$B_z$=($\phi_0/2\pi)z/\mid r \mid ^3$, using $z$=2.45$\mu$m
from the fit to the vortex in the (103) film. The ``vortex" in the (001) film
is resolution limited using our 4$\mu$m
SQUID pickup loop, but apparently has less then $\phi_0$ of flux.
The dashed curve is the prediction for a monopole source with
0.4$\phi_0$ total flux.

The source of the spontaneous moments in the $c$-axis
YBCO electrode is not known.
To our knowledge this is the first time they have been magnetically imaged
in $c$-axis
YBCO. Since there are no grain boundaries in the (001) films,
the magnetic roughness must be related to some other
aspect of the
morphology of the sample. We speculate that it is due to the presence of
defects in the film.
Figure 2 shows a comparison of SSM and scanning electron microscope (SEM)
images of the same region of $c$-axis
film. There appears to be some correlation between defects seen in the
SEM images and the magnetic structure. The ovals in Figure 2
are intended as guides to the eye of some of these correlations.
Since the predominant defects in these films are Y211 precipitates,
we speculate that these are causing the spontaneous magnetization
in these films. Y211 is insulating, with an antiferromagnetic
ordering temperature between 15 and 30K \cite{weidinger,chatt,meyer}

Figure 3 shows a series of scanning SQUID microscope images of one of
our samples for selected temperatures\cite{vartapl}. In these
images the sample was cooled in nominal zero field, and then imaged
while warming with a square pickup loop 17.8$\mu$m on a side. The
individual ``fractional" vortices are not resolved with this larger
pickup loop, as they were using the 4 $\mu$m pickup loop.
As the critical temperature ($\sim$ 80K for this film) is approached,
``telegraph noise" develops in the images, presumably due to reversals
in the sign of the flux generated at particular sites. This
``telegraph noise" appears to be driven by interactions with the SQUID
pickup loop, as evidenced by the streaking of the images, in the (vertical)
scan direction. The total variation in flux decreases as T$_c$ is
approached, until above T$_c$ the only magnetic signals present are a
few dipoles, presumably due to ferromagnetic particles in the sample.
These residual features are square, reflecting the shape of the pickup
loop. These images demonstrate that the spontaneous magnetization seen
in these samples is intimately related to the superconductivity, and
that the defects seen in these films are only weakly, if at all, ferromagnetic
above the superconducting critical temperature.

The spontaneous magnetization in these samples is qualitatively similar
to that seen\cite{mota} in granular BSCCO samples which show the
paramagnetic Meissner (Wohlleben) effect, both
in terms of its absolute magnitude and its random orientation.
However, there is an important
difference: these films show diamagnetic shielding at fields
of a few milliGauss.
An example is shown in Fig. 4. Fig. 4a is a scanning SQUID microscope image
of a sample cooled in an externally applied field of 2.5mG, using
a 17.8$\mu$m square pickup loop, and imaged in the same field at a sample
temperature of 60K. Also shown are histograms of the relative frequency
of flux observed in the upper (001) section of the YBCO film (Fig. 4(b)),
and the lower (103) YBCO (Fig. 4(c)). These histograms show that although
the measured flux above the (001) section has a large distribution width,
it has the
same average flux as above the (103) section. Both show
diamagnetic shielding on average.
In contrast, granular BSCCO samples show paramagnetic
shielding when cooled in fields below about 0.5G\cite{mota,sigrmp}.

Our measurements show that spontaneously generated moments can occur in samples
of high-T$_c$ superconductors
which do not have $\pi$-loops in the presence of grain boundaries.
One explanation for these observations is the existence of an out-of-phase
component to the dominant $d_{x^2-y^2}$ superconducting order parameter,
as predicted in the presence of interfaces or impurities \cite{matsumoto}.
Since such a complex order parameter would break time-reversal symmetry,
this interpretation is consistent with the presence of ``vortices" which
appear to have less than the superconducting quantum of flux. An alternate
explanation \cite{clemconv}
is the pinning of a vortex tangle, produced near T$_c$ in
a Kosterlitz-Thouless \cite{kosterlitz} type transition in the nearly two
dimensional superconductor YBCO. This pinning would be facilitated by
the disorder present in these films. This explanation seems inconsistent
with the temperature dependent measurements (Fig. 3), which show more
pronounced magnetic structure as T is reduced. However, it could be
argued that the vortex-antivortex fluctuations are too
rapid close to T$_c$ to be imaged with the SSM. Our measurements
are consistent with the absence of effects associated with
time-reversal symmetry breaking in
most measurements, and the appearance of such effects, which
may be associated with small concentrations of defects, in others.
Our experiments raise the possibility of intentionally introducing
time-reversal symmetry breaking effects by, for example, photolithographically
patterning small defects in high-T$_c$ samples. This could be applicable to
(e.g.) the fabrication of elements for quantum
computation \cite{geshkenbein}, opening new perspectives in the design of such devices
without necessarily using Josephson junctions. Possible advantages may be also related to the
smaller values of the magnetic flux associated with ``fractional vortices".

We would like to thank J.R. Clem, V.G. Kogan,
K.A. Moler and C.C. Tsuei for useful discussions,
M.B. Ketchen for the design, and M. Bhushan for the fabrication of
the SQUIDs used in this study, and F. Carillo, E. Sarnelli and G. Testa for
some help in
the preparation of the junctions. F.T. has been partially supported by the
projects PRA-INFM ``HTS Devices" and by a MURST COFIN98 program (Italy).

\begin{figure}
\caption{(a) Schematic of the geometry of a (001)/(103) biepitaxial
YBCO grain boundary sample. (b) Scanning SQUID microscope image of such a
grain boundary, cooled in $\sim$5mG field and imaged at 4.2K with an
octagonal pickup loop 4$\mu$m in diameter. The grain boundary, indicated
by a dashed line, runs nearly vertically through the center of the image.
(c) The dots are a
cross-section through the image (b) as indicated by the nearly horizontal
dotted line, through a ``vortex" on either side of the grain boundary.
The lines are fits to the data as described in the text.
}

\vspace{0.3in}
\caption{
Comparison between scanning SQUID microscope (SSM) and scanning electron microscope
(SEM) images of the same area of a
(001) YBCO thin film. The labelled regions are guides to the eye for areas
of correlation between the images.
}

\vspace{0.3in}
\caption{
Scanning SQUID microscope images of a (001)/(103) biepitaxial SQUID, cooled
in zero field, and imaged with a 17.8$\mu$m diameter square pickup loop, for
selected temperatures.
}

\vspace{0.3in}
\caption{(a)
Scanning SQUID microscope images of a (001)/(103) biepitaxial SQUID,
cooled in a field of 3mG, and imaged at T=60K. Also shown are histograms
of the frequency of occurence of SQUID sensor fluxes above areas (outlined
by dashed lines) in the (001) (b) and (103) (c) regions of the sample.
The (001) region shows a much broader distribution due to spontaneous
magnetization, but both regions show diamagnetic shielding.
}

\label{autonum}
\end{figure}

\end{multicols}

\begin{references}
\bibitem{kogprb} J.R. Kirtley, V.G. Kogan, J.R. Clem, and K.A. Moler,
Phys. Rev. B {\bf 59} 4343 (1999). The model used here is
for vortices oriented parallel to, and emerging from a
surface normal to, the planes. The (103) film surfaces are at a 45$^o$ angle to the
planes, so this model is not strictly applicable, and the
value for $\lambda_c$ reported here is only approximate.
\end{references}
\end{document}